\documentclass[aps,prc,reprint,superscriptaddress,nofootinbib]{revtex4-2}

\usepackage{amsmath}
\usepackage{amssymb}
\usepackage{graphicx}
\usepackage{bm}
\usepackage{braket}
\usepackage{slashed}

\begin{document}


\title{
Bootstrapping the deuteron
}



\author{Dong Bai}
\email{dbai@hhu.edu.cn}
\affiliation{College of Science, Hohai University, Nanjing 211100, Jiangsu, China}

%
%



\begin{abstract}

Bootstrap is a novel and ambitious paradigm for quantum physics.
It aims to solve the target problems by exploiting theoretical constraints from general physical principles and self-consistency conditions.
The bootstrap philosophy dates back to the 1960s.
Its real power has been recognized only recently in, e.g., conformal field theories and relativistic scattering amplitudes.
Inspired by [X.\ Han, S.\ A.\ Hartnoll, and J.\ Kruthoff, Phys.\ Rev.\ Lett.\ {\bf 125}, 041601 (2020)], 
we report the first bootstrap results in low-energy nuclear physics, where deuteron, with
its Hamiltonian given by pionless effective field theory in harmonic oscillator space, is solved by directly exploiting the most fundamental quantum mechanical requirement that probability should never be negative.
Our study shows that the bootstrap method can be helpful in studying realistic nuclear systems.

\end{abstract}


\maketitle

\section{Introduction}

Nonrelativistic quantum few- and many-body methods play a fundamental role in low-energy nuclear theory \cite{Gloeckle:1983,Ring:1980}.
It is always crucial to explore new methods and new paradigms.
Some noticeable progress has been made very recently in this direction.
Inspired by the huge success of artificial intelligence (AI),
new variational methods based on artificial neural networks (ANNs) are put forward for few-nucleon systems,
incorporating the latest advances in AI into conventional variational methods \cite{Keeble:2019bkv,Adams:2020aax,Gnech:2021wfn}.
Also, lots of interests are stimulated in developing new theoretical methods on quantum devices,
encouraged by the public accessibility of quantum computing clouds via the internet
\cite{Dumitrescu:2018njn,Klco:2018kyo,Roggero:2018hrn,Lee:2019zze,Roggero:2019myu,Lacroix:2020nhy,DiMatteo:2020dhe,Roggero:2020sgd,Cervia:2020fkk,RuizGuzman:2021cdj,Du:2021ctr,Siwach:2021tym,Stetcu:2021cbj,Baroni:2021xtl,Guzman:2021ede}.
These achievements enrich our tools to study nuclear systems.

Bootstrap is a novel and ambitious paradigm for quantum physics.
It was first advocated by Geoffrey Chew in the 1960s as a potential solution to the problem of the strong interaction,
under the guidance of the philosophical thinking that quantum physics problems could be solved solely by exploiting general physical principles and self-consistency 
conditions \cite{Chew:1962}.
Although influential temporarily,
the bootstrap program turns out to be over-ambitious and immature as a theory of the strong interaction. It is not competitive with quantum chromodynamics (QCD) discovered in the early 1970s and has been ignored largely by the community since then.
It gets revived only recently as alternative yet powerful approaches to conformal field theories and relativistic scattering amplitudes,
 resulting in new advances known as conformal bootstrap \cite{Poland:2018epd} and S-matrix bootstrap \cite{Paulos:2017fhb} in literature.
It is found that conformal bootstrap gives the most accurate predictions for critical exponents of three-dimensional Ising model \cite{El-Showk:2014dwa},
while the S-matrix bootstrap
provides valuable bounds on free parameters in relativistic scattering amplitudes \cite{Guerrieri:2021tak}. 
These bootstrap studies broaden our horizons significantly and reveal the hidden elegancy and beauty of relativistic quantum theories.
It is desirable to continue expanding the applicable scope of bootstrap.
Inspired by progress on string theory \cite{Anderson:2016rcw,Lin:2020mme}, Han \emph{et al.}~take the key step towards generalizing bootstrap from relativistic quantum theories to nonrelativistic quantum mechanics \cite{Han:2020bkb}.
They show that the quartic anharmonic oscillator can be solved by directly exploiting the most fundamental requirement in quantum mechanics, i.e., 
probability should never be negative.
Mathematically, this is done by requiring the so-called bootstrap matrix to be always positive semidefinite,
with its matrix elements obtained via some recursive relations (see Section \ref{BS} for details).
Their bootstrap method is soon applied to other quantum mechanical problems, 
such as harmonic oscillator, hydrogen, double-well potential, and Bloch bands \cite{Berenstein:2021dyf,Bhattacharya:2021btd,Aikawa:2021eai,Tchoumakov:2021mnh,Aikawa:2021qbl,Berenstein:2021loy,Du:2021hfw}.
Besides intelligential novelty, 
it is possible that these studies may eventually lead to an alternative formulation of quantum mechanics
that will be important in the future.
Historically, the evolution
from Newtonian mechanics to analytical mechanics,
lasting for several generations,
prepares the essential ground for the revolutionary advance from classical mechanics to quantum mechanics in the twentieth century.
For quantum mechanics,
although born one hundred years ago,
it may still have some new alternative formulations to be explored,
which play the role of the springboard to new physics revolutions.
Bootstrap is a candidate for such alternative formulations. 
Compared with some traditional methods, it is prized for the capability of making manifest the full power of fundamental physical principles and theoretical self-consistency.

As mentioned before, recursive relations are important in many previous quantum mechanical bootstrap studies,
thanks to which the bootstrap matrix elements 
could be determined up to a finite number of free parameters without solving Schr\"odinger equations.
Although elegant, the need for recursive relations restricts the form of potentials in Schr\"odinger equations to simple ones like polynomial and trigonometric potentials,
and thus narrows the applicable scope of quantum mechanical bootstrap method.
At present, quantum mechanical bootstrap method has not been used to study any realistic systems in low-energy nuclear physics.
Considering the complexity of nuclear forces, it is no longer easy to work out recursive relations for bootstrap matrix elements.
In this work, we formulate quantum mechanical bootstrap method in a more flexible form and report the first bootstrap result in low-energy nuclear physics.
The deuteron energy is calculated by exploiting the same self-consistency conditions as Ref.~\cite{Han:2020bkb}, but no recursive relations are used.
Deuteron is always a valuable test ground for new methods and new paradigms in low-energy nuclear physics. 
It has been adopted as the target system in several ANN and quantum computing studies \cite{Keeble:2019bkv,Dumitrescu:2018njn,Siwach:2021tym,Du:2021ctr,Adams:2020aax,Gnech:2021wfn}.
It is reasonable to start our bootstrap journey with deuteron as well.

The rest parts are organized as follows. 
In Section \ref{AQAO},
the  theoretical formalism is given for the quantum mechanical bootstrap approach to deuteron.
In Section \ref{Deuteron}, numerical results are given for the bootstrap calculations.
In Section \ref{Concl},
conclusions and remarks are given.

\section{
Theoretical Formalism
}
\label{AQAO}


\subsection{Bootstrap}
\label{BS}


In quantum mechanical bootstrap method, 
 the Schr\"odinger equation
$H\ket{E}=E\ket{E}$
is solved by
exploiting the self-consistency condition
\begin{align}
\braket{\Psi|\mathcal{O}^\dagger\mathcal{O}|\Psi}\geq 0,
\label{PC}
\end{align}
which holds for arbitrary states $\ket{\Psi}$
and arbitrary operators $\mathcal{O}$.
It results from the most fundamental requirement in quantum mechanics
that probability should never be negative, which is the essential prerequisite for Born's probabilistic interpretation.
To relate Eq.~\eqref{PC} to the target Schr\"odinger equation,
we take the state $\ket{\Psi}$ to be the eigenstate $\ket{E}$
and assume a general form of the operator 
$\mathcal{O}=\sum_{i=0}^K \alpha_i \mathcal{O}_i$,
with $\{\mathcal{O}_i\}$ being a set of operators
and $\{\alpha_i\}$ being the coefficients.
Eq.~\eqref{PC} then becomes
\begin{align}
\bm{\alpha}^\dagger\bm{\mathcal{M}}\,\bm{\alpha}\geq0,
\label{PC2}
\end{align}
with the vector $\bm{\alpha}=\{\alpha_0,\cdots,\alpha_K\}^\text{T}$.
$\bm{\mathcal{M}}$ of size $(K+1)\times(K+1)$ is known as the bootstrap matrix, with the matrix element given by $\mathcal{M}_{ij}=\braket{E|\mathcal{O}_i^\dagger\mathcal{O}_j|E}\equiv\braket{\mathcal{O}_i^\dagger\mathcal{O}_j}_E$.
Since $\bm{\alpha}$ could be chosen arbitrarily, 
the matrix $\bm{\mathcal{M}}$ should always be positive semidefinite, i.e.,
\begin{align}
\bm{\mathcal{M}}\succeq0.
\end{align}
Here, $\succeq$ is the standard mathematical notation for positive semidefiniteness.
If the bootstrap matrix $\bm{\mathcal{M}}$ is known in advance up to a few undetermined parameters such as the eigenenergy $E$, positive semi-definiteness could be used to carve out the allowable regions in the search space spanned by these parameters.
In some cases, the allowable regions may even shrink approximately to a number of points,
corresponding to different eigenstates of the target Hamiltonian.
In these cases, the Schr\"odinger equation is solved completely.

In Ref.~\cite{Han:2020bkb}, the quantum mechanical bootstrap method is used to study the quartic anharmonic oscillator
$H=p^2+x^2+gx^4$,
where
the operator $\mathcal{O}$ is taken to be
$\mathcal{O}=\sum_{i=0}^K\alpha_i x^i$.
The matrix elements of the bootstrap matrix $\bm{\mathcal{M}}$ are given by $\mathcal{M}_{ij}=\braket{x^{i+j}}_E$.
The recursive relation 
$4nE\braket{x^{n-1}}_E+n(n-1)(n-2)\braket{x^{n-3}}_E-4(n+1)\braket{x^{n+1}}_E-4g(n+2)\braket{x^{n+3}}_E=0$
is used to express all the matrix elements $\braket{x^n}_E$ with $n\geq4$ in terms of $E$
and $\braket{x^2}_E$.  
For odd $n$, $\braket{x^n}_E$ is equal to zero automatically as a result of the $Z_2$ symmetry.
With a given $K$, 
the allowable regions in the two-dimensional search space spanned by $E$ and $\braket{x^2}_E$ could be spotted
by the requirement that $\bm{\mathcal{M}}$ should be positive semidefinite.
As shown in Fig.~1 of Ref.~\cite{Han:2020bkb},
the sizes of the allowable regions decrease quickly as $K$ increases and shrink to a number of points approximately at large $K$s.
%
Similar strategies are also used in many other quantum mechanical bootstrap studies \cite{Berenstein:2021dyf,Bhattacharya:2021btd,Aikawa:2021eai,Tchoumakov:2021mnh,Aikawa:2021qbl,Berenstein:2021loy,Du:2021hfw}.
However, as we mention before, for general potentials such as those encountered in low-energy nuclear physics, it is not easy to work out useful recursive relations.

\subsection{Deuteron}

Deuteron is one of the simplest nuclei in nature. It could be viewed
as a two-body system of proton and neutron interacting with each other via meson exchanges. 
Its experimental energy is found to be $E_d^\text{exp}=-2.224566$ MeV \cite{Wang:2012},
resulting in a binding momentum $Q\approx46\ \text{MeV}$ much smaller than the mass of the lightest meson $M_\pi\approx138\ \text{MeV}$.
Therefore, for deuteron, one can simulate meson-exchange potentials by a series of contact interactions
with increasing numbers of derivatives.
When formulated in terms of modern effective field theory,
the above picture gives rise to pionless effective field theory ($\slashed{\pi}$EFT) \cite{Bedaque:2002mn,Hammer:2019poc},
where the relevant degrees of freedom are just nucleons and the breakdown scale is often estimated to be $\Lambda\approx M_\pi$.
Compared with many phenomenological nucleon-nucleon potentials, $\slashed{\pi}$EFT provides a systematically improvable model-independent
description of deuteron.
It is also used to study heavier nuclei with more nucleons, as well as some atomic systems \cite{Kirscher:2015ana,Bazak:2018qnu}.

At leading order (LO), the bare nucleon-nucleon potential from $\slashed{\pi}$EFT is given in \emph{free space} by
\begin{align}
V_\text{NN}^\text{LO}(\bm{p}',\bm{p})=C_S+C_T\,\bm{\sigma}_1\cdot\bm{\sigma}_2.
\end{align}
Here, $\bm{p}$ and $\bm{p}'$ are relative momenta in initial and final states,
$C_S$ and $C_T$ are low-energy constants, and $\bm{\sigma}_1$ and $\bm{\sigma}_2$ are the Pauli matrix vectors.
For deuteron, only the bare component from the ${}^{3}S_{1}$ channel is relevant
\begin{align}
V_\text{NN}^\text{LO}({}^{3}{S}_1)=\widetilde{C}_{{}^{3}S_1}\equiv C_S-3C_T.
\end{align}
It is well-known that the bare potential by itself would produce ultraviolet divergences in two-body calculations.
Therefore, it has to be regularized before taken as the input to numerical solvers of Schr\"odinger equations.
There are various choices for regulators, such as Gaussian and super-Gaussian regulators.
From the technical viewpoint, these regularized nucleon-nucleon potentials are not friendly to the quantum mechanical bootstrap approach.
It is no longer easy to work out recursive relations for bootstrap matrix elements,
making it less straightforward to apply the method in Ref.~\cite{Han:2020bkb} to deuteron.

In this work, we adopt the deuteron Hamiltonian from $\slashed{\pi}$EFT in \emph{harmonic oscillator space} (aka oscillator $\slashed{\pi}$EFT)  instead of free space \cite{Dumitrescu:2018njn,Binder:2015trg,Bansal:2017pwn}
\begin{align}
&H_{N_\text{max}}=\sum_{i,j=0}^{N_\text{max}}H_{ij}\ket{i}\bra{j}
\equiv\sum_{i,j=0}^{N_\text{max}}(T_{ij}+V_{ij})\ket{i}\bra{j},
\label{HamiD}\\
&T_{ij}=\braket{i|T|j}\nonumber\\
&\ \ \ \,=\frac{\hbar\omega}{2}\big[(2j+3/2)\delta_{ij}-\sqrt{j(j+1/2)}\delta_{i+1,j}\nonumber\\
&\ \ \ \,-\sqrt{(j+1)(j+3/2)}\delta_{i-1,j}\big],\\
&V_{ij}=\braket{i|V|j}=V_0\delta_{j0}\delta_{ij}.
\end{align}
Here, $\ket{i}$ is the $S$-wave eigenstate of harmonic oscillator with $i$ nodes in the radial wave function,
$T_{ij}$ and $V_{ij}$ are matrix elements for kinetic and potential energies,
$H_{ij}$ is the matrix element for the total Hamiltonian,
and $\hbar\omega=7$ MeV is the harmonic oscillator frequency.
It is easy to see that the Hamiltonian matrix, as a band matrix, is real and symmetric,
i.e., $H_{ij}^*=H_{ij}$ and $H_{ij}=H_{ji}$. 
Loosely speaking, harmonic oscillator space provides a specific discretization of 
free space, and oscillator effective field theory is a novel alternative to 
lattice effective field theory which discretizes free space by a lattice \cite{Binder:2015trg,Bansal:2017pwn,Yang:2016brl}.
With $V_0=-5.68658111$ MeV,
the deuteron energy is found to be $E_d^\text{th}=-2.221$ MeV in the limit of $N_\text{max}\to\infty$, numerically close to the experimental one.
This shows that the Hamiltonian in Eq.~\eqref{HamiD} is indeed reliable for deuteron.
Noticeably, this Hamiltonian has been adopted in several publications to explore the quantum-computing approach to 
low-energy nuclear physics \cite{Dumitrescu:2018njn,Siwach:2021tym}. 
It could also be a good starting point to develop the quantum mechanical bootstrap method to
solve nuclear physics problems.

\begin{widetext}

\subsection{Bootstrap for Deuteron}

To bootstrap the deuteron, we consider the following operator
\begin{align}
\mathcal{O}&=\sum_{i\leq j}\ket{i}\bra{j}
\equiv\sum_{i\leq j}\mathcal{O}_{ij}\nonumber\\
&=\mathcal{O}_{00}+\cdots+\mathcal{O}_{0,N_\text{max}}
+\mathcal{O}_{11}+\cdots+\mathcal{O}_{1,N_\text{max}}\nonumber\\
&+\cdots+\mathcal{O}_{N_\text{max}-1,N_\text{max}-1}+\mathcal{O}_{N_\text{max}-1,N_\text{max}}+\mathcal{O}_{N_\text{max},N_\text{max}},\end{align}
where the harmonic oscillator space is truncated by $N_\text{max}$,
and $i,j=0,\cdots,N_\text{max}$.
The operators $\{\mathcal{O}_{ij}\}$ satisfy $\mathcal{O}_{ij}^\dagger=\mathcal{O}_{ji}$, 
$\mathcal{O}_{ij}^\dagger\mathcal{O}_{kl}=\delta_{ik}\mathcal{O}_{jl}$,
and $\sum_{i=0}^{N_\text{max}}\mathcal{O}_{ii}=1$.
In total, we have $N_{\bm{\mathcal{M}}}\equiv (N_\text{max}+1)(N_\text{max}+2)/2$ elements in $\{\mathcal{O}_{ij}\}$,
from which we can construct
the corresponding bootstrap matrix of size $N_{\bm{\mathcal{M}}}\times N_{\bm{\mathcal{M}}}$.
Some examples of bootstrap matrices are given by
\begin{align}
\bm{\mathcal{M}}_2=\left(
\begin{array}{cccccc}
\braket{\mathcal{O}_{00}}_{E_2} & \braket{\mathcal{O}_{01}}_{E_2} & \braket{\mathcal{O}_{02}}_{E_2} & 0 & 0 & 0 \\
\braket{\mathcal{O}_{01}}_{E_2} & \braket{\mathcal{O}_{11}}_{E_2} & \braket{\mathcal{O}_{12}}_{E_2} & 0 & 0 & 0 \\
\braket{\mathcal{O}_{02}}_{E_2} & \braket{\mathcal{O}_{12}}_{E_2} & \braket{\mathcal{O}_{22}}_{E_2} & 0 & 0 & 0 \\
0 & 0 & 0 & \braket{\mathcal{O}_{11}}_{E_2} & \braket{\mathcal{O}_{12}}_{E_2} & 0 \\
0 & 0 & 0 & \braket{\mathcal{O}_{12}}_{E_2} & \braket{\mathcal{O}_{22}}_{E_2} & 0 \\
0 & 0 & 0 & 0 & 0 & \braket{\mathcal{O}_{22}}_{E_2}
\end{array}
\right),
\label{BM3}
\end{align}

\begin{align}
\bm{\mathcal{M}}_3=\left(
\begin{array}{cccccccccc}
\braket{\mathcal{O}_{00}}_{E_3} & \braket{\mathcal{O}_{01}}_{E_3} & \braket{\mathcal{O}_{02}}_{E_3} & \braket{\mathcal{O}_{03}}_{E_3} & 0 & 0 & 0 & 0 & 0 & 0 \\
\braket{\mathcal{O}_{01}}_{E_3} & \braket{\mathcal{O}_{11}}_{E_3} & \braket{\mathcal{O}_{12}}_{E_3} & \braket{\mathcal{O}_{13}}_{E_3} & 0 & 0 & 0 & 0 & 0 & 0 \\
\braket{\mathcal{O}_{02}}_{E_3} & \braket{\mathcal{O}_{12}}_{E_3} & \braket{\mathcal{O}_{22}}_{E_3} & \braket{\mathcal{O}_{23}}_{E_3} & 0 & 0 & 0 & 0 & 0 & 0  \\
\braket{\mathcal{O}_{03}}_{E_3} & \braket{\mathcal{O}_{13}}_{E_3} & \braket{\mathcal{O}_{23}}_{E_3} & \braket{\mathcal{O}_{33}}_{E_3} & 0 & 0 & 0 & 0 & 0 & 0  \\
0 & 0 & 0 & 0 & \braket{\mathcal{O}_{11}}_{E_3} & \braket{\mathcal{O}_{12}}_{E_3} & \braket{\mathcal{O}_{13}}_{E_3} & 0 & 0 & 0 \\
0 & 0 & 0 & 0 & \braket{\mathcal{O}_{12}}_{E_3} & \braket{\mathcal{O}_{22}}_{E_3} & \braket{\mathcal{O}_{23}}_{E_3} & 0 & 0 & 0 \\
0 & 0 & 0 & 0 & \braket{\mathcal{O}_{13}}_{E_3} & \braket{\mathcal{O}_{23}}_{E_3} & \braket{\mathcal{O}_{33}}_{E_3} & 0 & 0 & 0 \\
0 & 0 & 0 & 0 & 0 & 0 & 0 & \braket{\mathcal{O}_{22}}_{E_3} & \braket{\mathcal{O}_{23}}_{E_3} & 0 \\
0 & 0 & & 0 & 0 & 0 & 0 & \braket{\mathcal{O}_{23}}_{E_3} & \braket{\mathcal{O}_{33}}_{E_3} & 0 \\
0 & 0 & 0 & 0 & 0 & 0 & 0 & 0 & 0 &  \braket{\mathcal{O}_{33}}_{E_3}
\end{array}
\right).
\label{BM4}
\end{align}
 Here, the subscripts in $\bm{\mathcal{M}}_2$ and $\bm{\mathcal{M}}_3$
emphasize that these are bootstrap matrices for $N_\text{max}=2$ and $N_\text{max}=3$,
while the subscripts $E_{2}$ and $E_3$ in each bootstrap matrix element are the eigenenergies of the corresponding deuteron Hamiltonians $H_{2}$ and $H_3$.
These notations can be easily generalized to $N_\text{max}>3$ and will be used in the following parts.
In Eqs.~\eqref{BM3} and \eqref{BM4},
the relation $\braket{\mathcal{O}_{ij}}_{E_{N_\text{max}}}=\braket{\mathcal{O}_{ji}}_{E_{N_\text{max}}}$ is used to do the simplification.
This relation could be understood from the fact that $\{\braket{i|E_{N_\text{max}}}\}$ are real,
i.e., $\braket{i|E_{N_\text{max}}}^\dagger=\braket{i|E_{N_\text{max}}}$.
The bootstrap matrices $\{\bm{\mathcal{M}}_{N_\text{max}}\}$ with $N_\text{max}>3$ also have similar forms.
These bootstrap matrices should always be positive semidefinite, i.e., $\bm{\mathcal{M}}_{N_\text{max}}\succeq0$, for the sake of self-consistency of quantum mechanics.
 
To bootstrap the deuteron, one needs to determine the numerical values of all the bootstrap matrix elements $\{\braket{\mathcal{O}_{ij}}_{E_{N_\text{max}}}\}$
at a given energy $E_{N_\text{max}}$.
In Ref.~\cite{Han:2020bkb}, bootstrap matrix elements for the quartic anharmonic oscillator are derived from recursive relations.
However, it is not easy to apply this method to deuteron. 
Instead, we make use of the following equations
\begin{align}
&\sum_{i=0}^{N_\text{max}} \braket{\mathcal{O}_{ii}}_{E_{N_\text{max}}}=1,\label{FBE}\\
&[ij]_{N_\text{max}}\equiv\left(\sum_{k=0}^{N_\text{max}}H_{(ki)}\braket{\mathcal{O}_{(kj)}}_{E_{N_\text{max}}}-E_{N_\text{max}}\braket{\mathcal{O}_{ij}}_{E_{N_\text{max}}}=0\right),\qquad\text{for $i\leq j$},
\label{SBE}
\end{align} 
with $i,j,k=0,\cdots,N_\text{max}$
and $(i,j)=(\text{min}(i,j),\text{max}(i,j))$.
The first equation corresponds to the completeness relation in harmonic oscillator space truncated by $N_\text{max}$,
while the second equation is derived from the self-consistency condition $\braket{H_{N_\text{max}}\mathcal{O}_{ij}}_{E_{N_\text{max}}}=E_{N_\text{max}}\braket{\mathcal{O}_{ij}}_{E_{N_\text{max}}}$.
The symmetric properties of $H_{ij}$ and $\braket{\mathcal{O}_{ij}}_{E_{N_\text{max}}}$ are also used in Eq.~\eqref{SBE},
allowing us to sort the subscripts $ij$ to their standard orders $(ij)$.
The basic idea is to impose as many constraints in the form of Eqs.~\eqref{FBE} and \eqref{SBE} as possible to determine the numerical values of $\{\braket{\mathcal{O}_{ij}}_{E_{N_\text{max}}}\}$ at a given energy $E$.
In principle, 
there could be totally $1+(N_\text{max}+1)(N_\text{max}+2)/2$ constraints,
making $\{\braket{\mathcal{O}_{ij}}_{E_{N_\text{max}}}\}$ over-constrained. 
As a result, one constraint in the form of Eq.~\eqref{SBE} has to be removed.
Interestingly, it turns out that which $[ij]_{N_\text{max}}$ is excluded does make a difference in the bootstrap calculations.

\section{Numerical Results}
\label{Deuteron}

In this section, we give the numerical results of our bootstrap approach to deuteron.
Four different $N_\text{max}$ truncations are considered for harmonic oscillator space,
including $N_\text{max}=2$, $3$, $9$, and $19$.
This
allows us to explore the quantum mechanical bootstrap method in a systematic way.
Compared with $N_\text{max}=9$ and 19,
the bootstrap calculations for $N_\text{max}=2$ and 3
are technically much simpler,
which also allows us to observe closely how the bootstrap method works.

\subsection{$N_\text{max}=2$}
\label{Nmax2}

At $N_\text{max}=2$, the deuteron Hamiltonian is given by
\begin{align}
H_2=\left(
\begin{array}{ccc}
-0.436581 & -4.28661 & 0 \\
-4.28661 & 12.25 & -7.82624 \\
0 & -7.82624 & 19.25
\end{array}
\right),
\end{align}
with the exact lowest eigenvalue found to be $E_2^{\text{exact}}=-2.045671$ MeV.
The corresponding bootstrap matrix is given by Eq.~\eqref{BM3}. For a given eigenenergy $E_2$,
the bootstrap matrix elements could be obtained by solving the following equations
\begin{align}
&\braket{\mathcal{O}_{00}}_{E_2}+\braket{\mathcal{O}_{11}}_{E_2}+\braket{\mathcal{O}_{22}}_{E_2}=1,\nonumber\\
&[00]_2\equiv\left(H_{00}\braket{\mathcal{O}_{00}}_{E_2}+H_{01}\braket{\mathcal{O}_{01}}_{E_2}+H_{02}\braket{\mathcal{O}_{02}}_{E_2}-E_2\braket{\mathcal{O}_{00}}_{E_2}=0\right),\nonumber\\
&[01]_2\equiv\left(H_{00}\braket{\mathcal{O}_{01}}_{E_2}+H_{01}\braket{\mathcal{O}_{11}}_{E_2}+H_{02}\braket{\mathcal{O}_{12}}_{E_2}-E_2\braket{\mathcal{O}_{01}}_{E_2}=0\right),\nonumber\\
&[11]_2\equiv\left(H_{01}\braket{\mathcal{O}_{01}}_{E_2}+H_{11}\braket{\mathcal{O}_{11}}_{E_2}+H_{12}\braket{\mathcal{O}_{12}}_{E_2}-E_2\braket{\mathcal{O}_{11}}_{E_2}=0\right),\nonumber\\
&[12]_2\equiv\left(H_{01}\braket{\mathcal{O}_{02}}_{E_2}+H_{11}\braket{\mathcal{O}_{12}}_{E_2}+H_{12}\braket{\mathcal{O}_{22}}_{E_2}-E_2\braket{\mathcal{O}_{12}}_{E_2}=0\right),\nonumber\\
&[22]_2\equiv\left(H_{02}\braket{\mathcal{O}_{02}}_{E_2}+H_{12}\braket{\mathcal{O}_{12}}_{E_2}+H_{22}\braket{\mathcal{O}_{22}}_{E_2}-E_2\braket{\mathcal{O}_{22}}_{E_2}=0\right).
\label{ConstN2}
\end{align}
The above equations could also be abbreviated to $\overline{[02]}_2$,
meaning that $[02]_2$ is excluded from the bootstrap constraints.
This notation can be easily generalized to $N_\text{max}>2$.
In this way, the bootstrap matrix $\bm{\mathcal{M}}_2$
is determined numerically for a given eigenenergy $E_2$.
The positive semidefinite property of $\bm{\mathcal{M}}_2$ 
could then be decided by checking, e.g., whether the eigenvalues $\{m_{2i}\}$ of the bootstrap matrix $\bm{\mathcal{M}}_2$ are all nonnegative. 
The realistic situation could be a bit complicated, as there could be negative eigenvalues with extremely tiny magnitudes close to zero.
To what extent these tiny negative eigenvalues can be safely regarded as zero is often crucial for determining the positive semidefinite property of the bootstrap matrix. 
A tolerance factor $\lambda$ is introduced in numerical calculations such that 
all the eigenvalues satisfying $|m_{2i}|\leq \lambda\, \text{max}(\{m_{2j}\})$
are regarded as zero for the bootstrap matrix $\bm{\mathcal{M}}_2$.
In \texttt{Mathematica}, there is a command called \texttt{PositiveSemidefiniteMatrixQ} for this purpose.
For $\lambda=10^{-2}$, $\bm{\mathcal{M}}_2$ is positive semidefinite at $-2.138857\ \text{MeV}\leq E_2\leq-1.610264\ \text{MeV}$.
For $\lambda=10^{-3}$, $\bm{\mathcal{M}}_2$ is positive semidefinite at $-2.057992\ \text{MeV}\leq E_2\leq-1.732635\ \text{MeV}$.
For $\lambda=10^{-4}$, $\bm{\mathcal{M}}_2$ is positive semidefinite at $-2.046958\ \text{MeV}\leq E_2\leq-2.036574\ \text{MeV}$.
For $\lambda=10^{-5}$, $\bm{\mathcal{M}}_2$ is positive semidefinite at $-2.045800\ \text{MeV}\leq E_2\leq-2.044787\ \text{MeV}$.
For $\lambda=10^{-6}$, $\bm{\mathcal{M}}_2$ is positive semidefinite at $-2.045683\ \text{MeV}\leq E_2\leq-2.045583\ \text{MeV}$.
For $\lambda=10^{-7}$, $\bm{\mathcal{M}}_2$ is positive semidefinite at $-2.045672\ \text{MeV}\leq E_2\leq-2.045663\ \text{MeV}$.
For $\lambda=10^{-8}$, $\bm{\mathcal{M}}_2$ is positive semidefinite at $-2.045671\ \text{MeV}\leq E_2\leq-2.0456701\ \text{MeV}$.
These results are also shown in Fig.~\ref{FigBS}(a).
It is straightforward to see that, as $\lambda$ decreases from $10^{-2}$ to $10^{-8}$,
the size of the allowed $E_2$ space decreases by about six orders of magnitude.
At $\lambda\leq 10^{-8}$,
the allowed $E_2$ space becomes so tiny that it practically determines the lowest eigenvalue of the deuteron Hamiltonian $H_2$,
giving results agreeing well with the exact solution $E^\text{exact}_2$ from direct matrix diagonalization.
It is interesting to note that, at a given $\lambda$, $E_2$ is bounded not only from above but also from below.
This is a novel feature of quantum mechanical bootstrap method compared to, e.g., the traditional variational method,
which generally gives only the upper bound of eigenvalues.
Last but not least, we would like to stress that, besides the deuteron ground state, the deuteron Hamiltonian $H_2$ has other eigenstates corresponding to continuum states of the proton-neutron system.
 The allowed $E_2$ space mentioned before is associated with the deuteron ground state only,
 and quantum mechanical bootstrap method can also be used to determine other eigenvalues of $H_2$.
 For example, the second lowest eigenenergy of $H_2$ is ${E'_2}^{\text{exact}}=8.5615817$ MeV.
 At $\lambda=10^{-8}$, quantum mechanical bootstrap method gives the result $8.5615817\ \text{MeV}\leq E_2'\leq8.5615818\ \text{MeV}$,
 in excellent agreement with the exact solution.
 In the rest subsections, we will focus on the deuteron ground state only,
 which is the primary concern of this work.

\subsection{$N_\text{max}=3$}

At $N_\text{max}=3$, the deuteron Hamiltonian is given by
\begin{align}
H_3=\left(
\begin{array}{cccc}
-0.436581 & -4.28661 & 0 & 0 \\
-4.28661 & 12.25 & -7.82624 & 0 \\
0 & -7.82624 & 19.25 & -11.3413 \\
0 & 0 & -11.3413 & 26.25
\end{array}
\right),
\end{align}
with the exact lowest eigenvalue found to be $E_3^\text{exact}=-2.143981$ MeV. 
The corresponding bootstrap matrix is given by Eq.~\eqref{BM4}.
For a given eigenenergy $E_3$,
the bootstrap matrix elements could be obtained from
\begin{align}
&\braket{\mathcal{O}_{00}}_{E_3}+\braket{\mathcal{O}_{11}}_{E_3}+\braket{\mathcal{O}_{22}}_{E_3}+\braket{\mathcal{O}_{33}}_{E_3}=1,\nonumber\\
&[00]_3\equiv \left(H_{00}\braket{\mathcal{O}_{00}}_{E_3}+H_{01}\braket{\mathcal{O}_{01}}_{E_3}+H_{02}\braket{\mathcal{O}_{02}}_{E_3}+H_{03}\braket{\mathcal{O}_{03}}_{E_3}-E_3\braket{\mathcal{O}_{00}}_{E_3}=0\right),\nonumber\\
&[01]_3\equiv \left(H_{00}\braket{\mathcal{O}_{01}}_{E_3}+H_{01}\braket{\mathcal{O}_{11}}_{E_3}+H_{02}\braket{\mathcal{O}_{12}}_{E_3}+H_{03}\braket{\mathcal{O}_{13}}_{E_3}-E_3\braket{\mathcal{O}_{01}}_{E_3}=0\right),\nonumber\\
&[02]_3\equiv \left(H_{00}\braket{\mathcal{O}_{02}}_{E_3}+H_{01}\braket{\mathcal{O}_{12}}_{E_3}+H_{02}\braket{\mathcal{O}_{22}}_{E_3}+H_{03}\braket{\mathcal{O}_{23}}_{E_3}-E_3\braket{\mathcal{O}_{02}}_{E_3}=0\right),\nonumber\\
&[03]_3\equiv \left(H_{00}\braket{\mathcal{O}_{03}}_{E_3}+H_{01}\braket{\mathcal{O}_{13}}_{E_3}+H_{02}\braket{\mathcal{O}_{23}}_{E_3}+H_{03}\braket{\mathcal{O}_{33}}_{E_3}-E_3\braket{\mathcal{O}_{03}}_{E_3}=0\right),\nonumber\\
&[11]_3\equiv \left(H_{01}\braket{\mathcal{O}_{01}}_{E_3}+H_{11}\braket{\mathcal{O}_{11}}_{E_3}+H_{12}\braket{\mathcal{O}_{12}}_{E_3}+H_{13}\braket{\mathcal{O}_{13}}_{E_3}-E_3\braket{\mathcal{O}_{11}}_{E_3}=0\right),\nonumber\\
&[12]_3\equiv \left(H_{01}\braket{\mathcal{O}_{02}}_{E_3}+H_{11}\braket{\mathcal{O}_{12}}_{E_3}+H_{12}\braket{\mathcal{O}_{22}}_{E_3}+H_{13}\braket{\mathcal{O}_{23}}_{E_3}-E_3\braket{\mathcal{O}_{12}}_{E_3}=0\right),\nonumber\\
&[22]_3\equiv \left(H_{02}\braket{\mathcal{O}_{02}}_{E_3}+H_{12}\braket{\mathcal{O}_{12}}_{E_3}+H_{22}\braket{\mathcal{O}_{22}}_{E_3}+H_{23}\braket{\mathcal{O}_{23}}_{E_3}-E_3\braket{\mathcal{O}_{22}}_{E_3}=0\right),\nonumber\\
&[23]_3\equiv \left(H_{02}\braket{\mathcal{O}_{03}}_{E_3}+H_{12}\braket{\mathcal{O}_{13}}_{E_3}+H_{22}\braket{\mathcal{O}_{23}}_{E_3}+H_{23}\braket{\mathcal{O}_{33}}_{E_3}-E_3\braket{\mathcal{O}_{23}}_{E_3}=0\right),\nonumber\\
&[33]_3\equiv \left(H_{03}\braket{\mathcal{O}_{03}}_{E_3}+H_{13}\braket{\mathcal{O}_{13}}_{E_3}+H_{23}\braket{\mathcal{O}_{23}}_{E_3}+H_{33}\braket{\mathcal{O}_{33}}_{E_3}-E_3\braket{\mathcal{O}_{33}}_{E_3}=0\right).
\end{align}
Following the conventions of Section \ref{Nmax2}, the above equations are abbreviated to $\overline{[13]}_3$,
which means that $[13]_3$ is not included in the bootstrap constraints.
For $\lambda=10^{-2}$, the bootstrap matrix $\bm{\mathcal{M}}_3$ is positive semidefinite at $-2.220264\ \text{MeV}\leq E_3\leq-1.6758798\ \text{MeV}$. 
For $\lambda=10^{-3}$, $\bm{\mathcal{M}}_3$ is positive semidefinite at $-2.157756\ \text{MeV}\leq E_3\leq-2.024596\ \text{MeV}$. 
For $\lambda=10^{-4}$, $\bm{\mathcal{M}}_3$ is positive semidefinite at $-2.1455807\ \text{MeV}\leq E_3\leq-2.130422\ \text{MeV}$. 
For $\lambda=10^{-5}$, $\bm{\mathcal{M}}_3$ is positive semidefinite at $-2.144144\ \text{MeV}\leq E_3\leq-2.142806\ \text{MeV}$. 
For $\lambda=10^{-6}$, $\bm{\mathcal{M}}_3$ is positive semidefinite at $-2.143997\ \text{MeV}\leq E_3\leq-2.143865\ \text{MeV}$. 
For $\lambda=10^{-7}$, $\bm{\mathcal{M}}_3$ is positive semidefinite at $-2.143982\ \text{MeV}\leq E_3\leq-2.1439695\ \text{MeV}$. 
For $\lambda=10^{-8}$, $\bm{\mathcal{M}}_3$ is positive semidefinite at $-2.143981\ \text{MeV}\leq E_3\leq-2.143980\ \text{MeV}$,
which is in excellent agreement with the exact result $E_3^\text{exact}=-2.143981$ MeV. 
These results are also shown in Fig.~\ref{FigBS}(b),
which is similar to the results of $N_\text{max}=2$ in Fig.~\ref{FigBS}(a).

\begin{figure*}


  \includegraphics[width=0.495\linewidth]{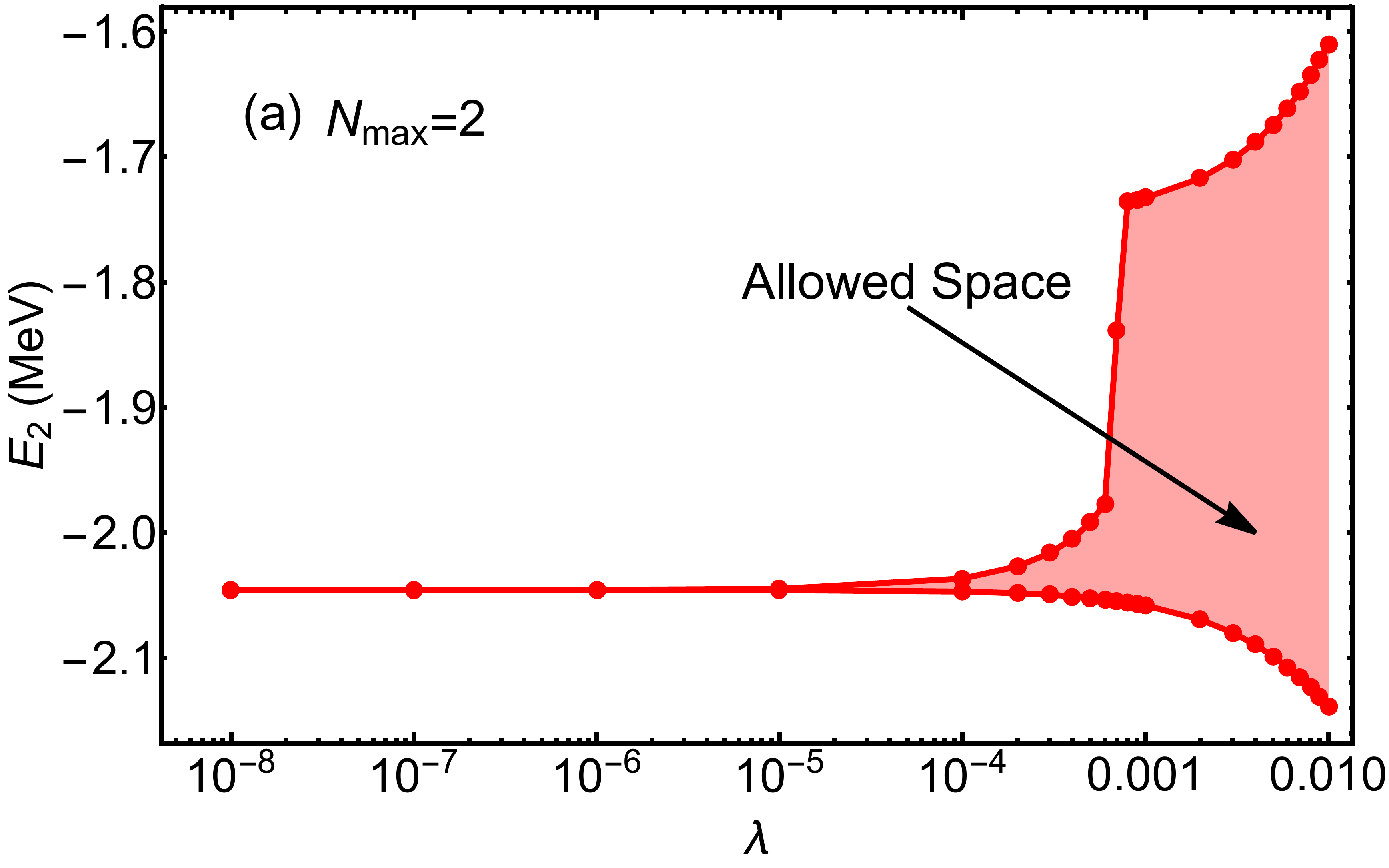}
  \includegraphics[width=0.495\linewidth]{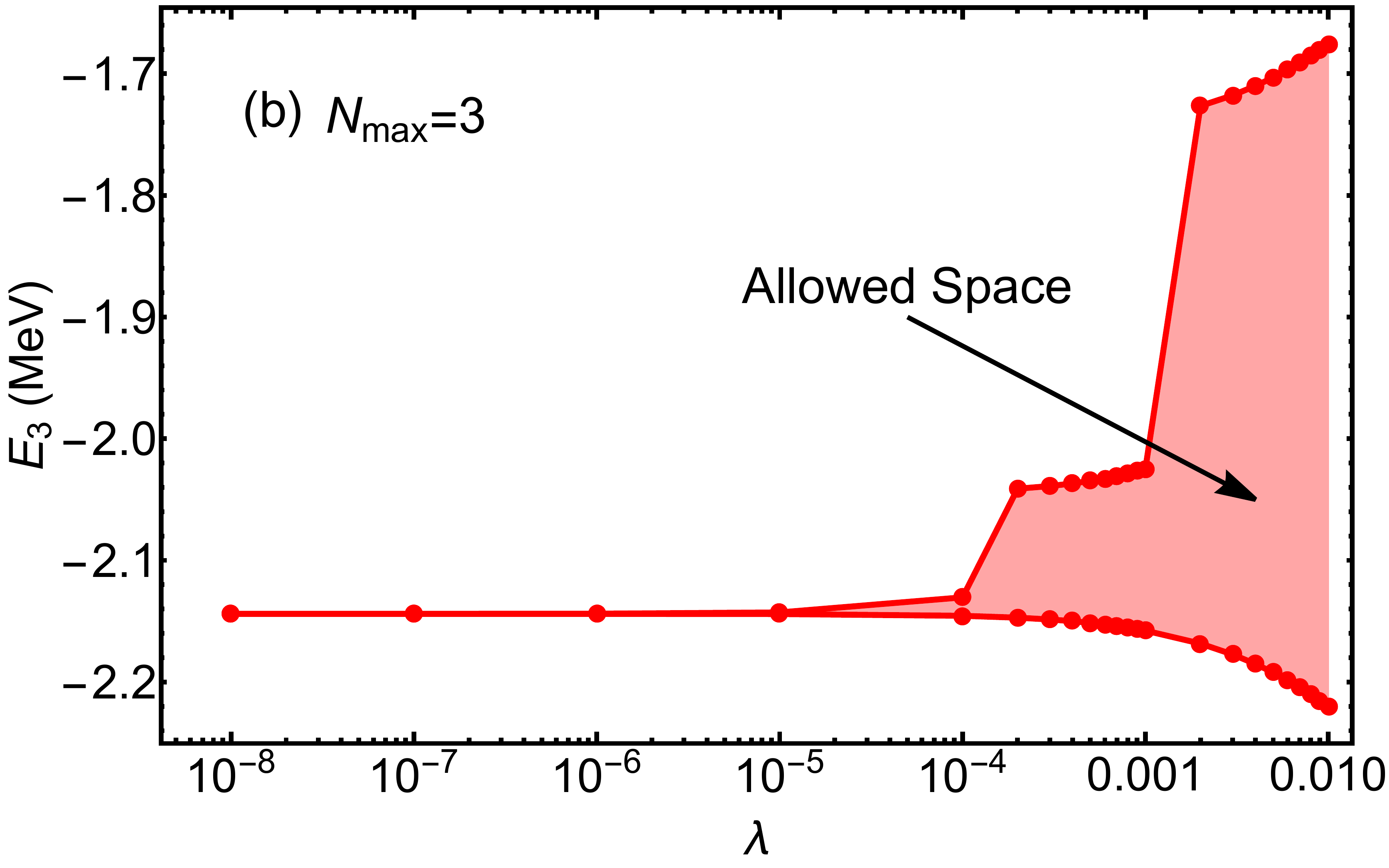}
    
  \includegraphics[width=0.495\linewidth]{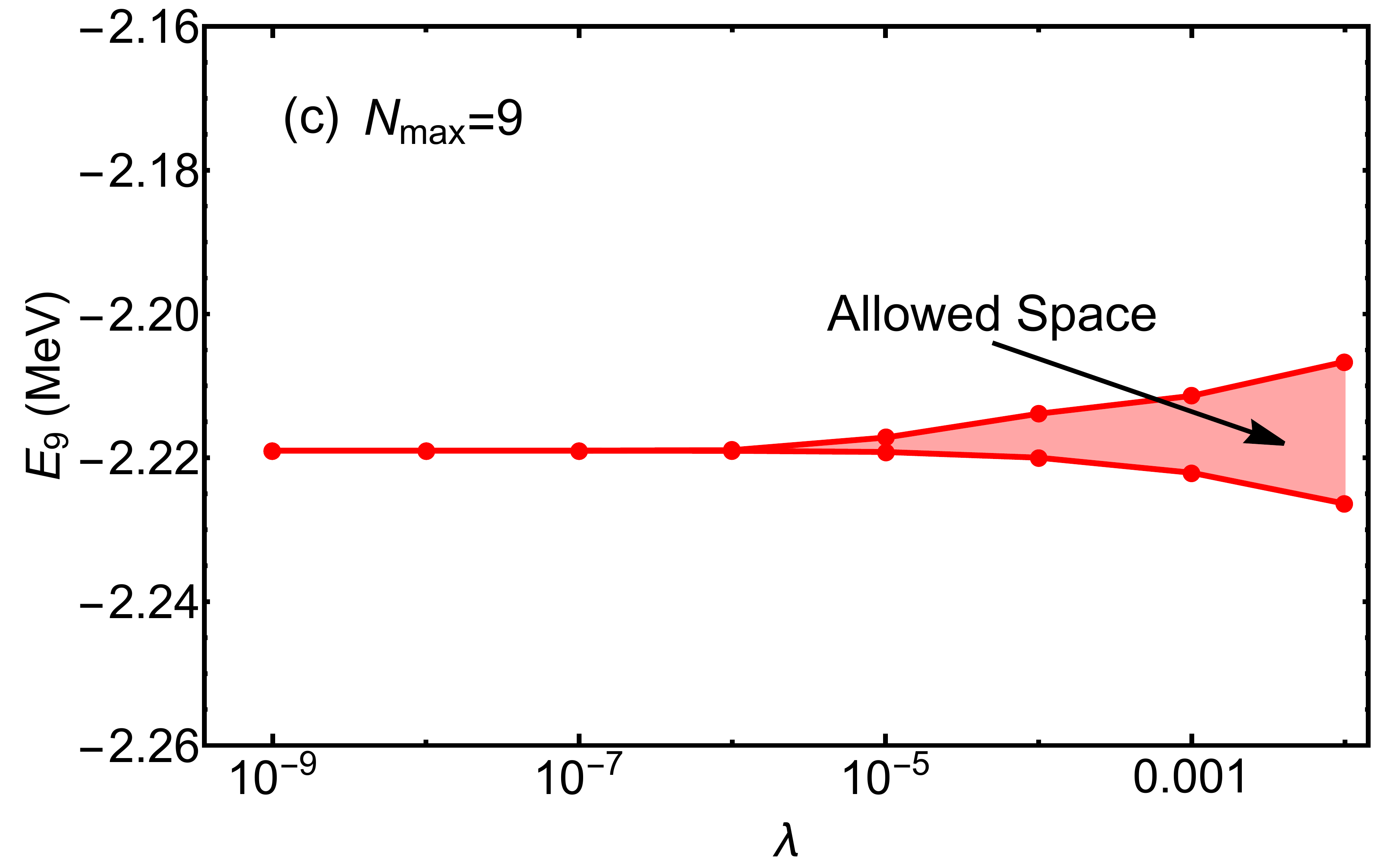}    
  \includegraphics[width=0.495\linewidth]{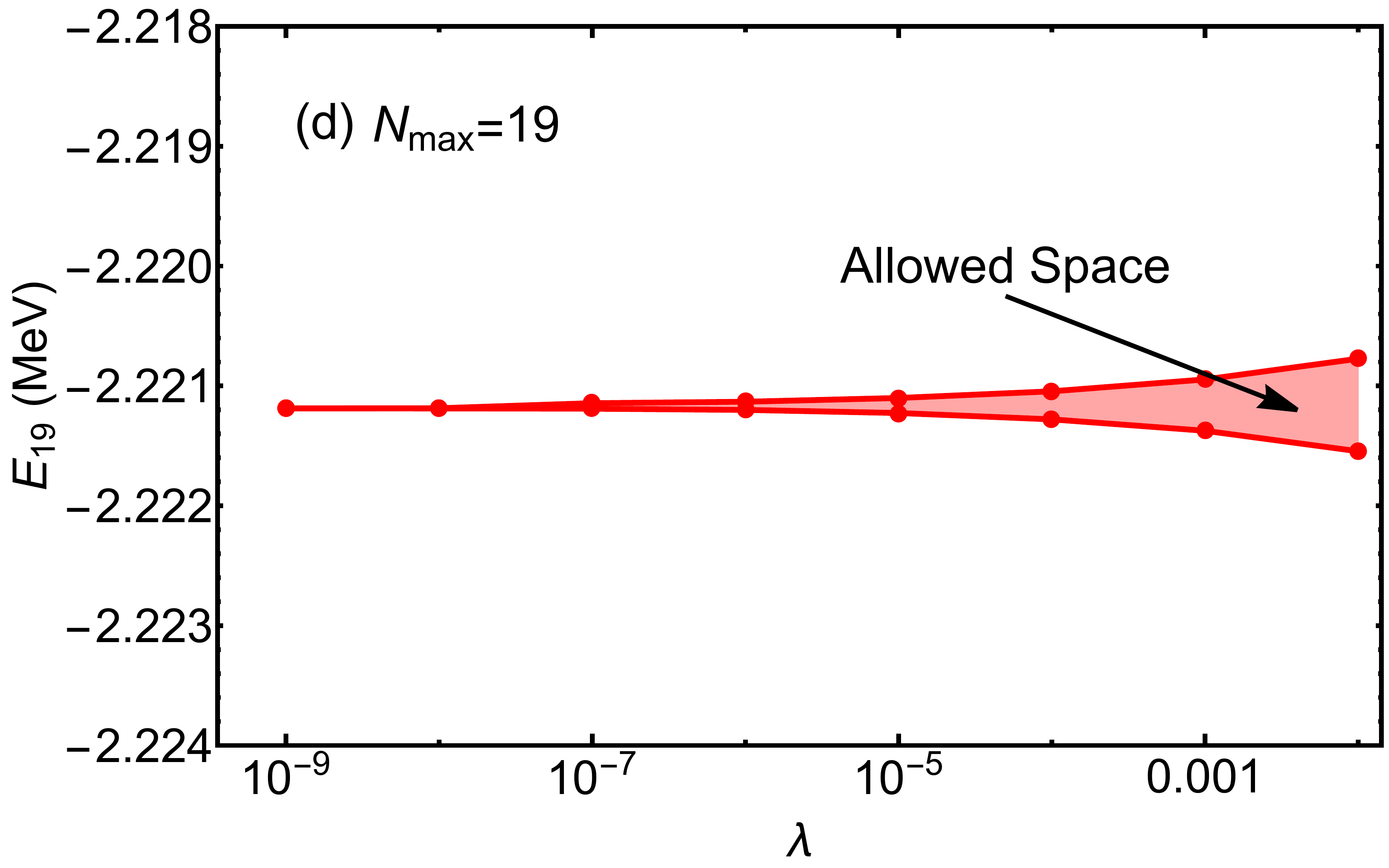}

  
  \caption{
The allowed $E_{N_\text{max}}$ spaces (light red) for the lowest eigenvalues of the deuteron Hamiltonians $H_{N_\text{max}}$ with respect to  tolerance factors $\lambda$.
Four different truncations are considered for the harmonic oscillator space with
(a) $N_\text{max}=2$,
(b) $N_\text{max}=3$, (c) $N_\text{max}=9$, and (d) $N_\text{max}=19$. 
The bootstrap constraints $\overline{[02]}_2$, $\overline{[13]}_3$, $\overline{[09]}_9$,
and $\overline{[0,19]}_{19}$ are imposed to bootstrap the deuteron 
at different $N_\text{max}$s.
The red points are the mesh points used to plot the upper and lower boundaries of the allowed $E_{N_\text{max}}$ spaces.
      }
  \label{FigBS}
  
  
\end{figure*}

\end{widetext}

\subsection{$N_\text{max}=9$ and $N_\text{max}=19$}

At $N_\text{max}=9$ and $N_\text{max}=19$, the deuteron Hamiltonians are given by real symmetric matrices of sizes $10\times10$ and $20\times20$.
The exact lowest eigenvalues are found to be $E^\text{exact}_9=-2.219002$ MeV and $E^\text{exact}_{19}=-2.221187$ MeV, respectively.
The corresponding bootstrap matrices are of sizes $55\times55$ and $210\times210$.
For simplicity, the explicit forms of these matrices are not given here, which are a bit lengthy compared to the bootstrap matrices at $N_\text{max}=2$ and 3.
To bootstrap the lowest eigenvalues, the bootstrap constraints $\overline{[09]}_9$ and $\overline{[0,19]}_{19}$
are used for $N_\text{max}=9$ and $N_\text{max}=19$.
These two sets of constraints have 55 and 210 equations, respectively.
Therefore, for given values of $E_9$ and $E_{19}$,
the bootstrap matrix elements $\{\braket{\mathcal{O}_{ij}}_{E_9}\}$ and $\{\braket{\mathcal{O}_{ij}}_{E_{19}}\}$
can be determined from these constraints.
In Figs.~\ref{FigBS}(c) and \ref{FigBS}(d), 
we plot the allowed $E_9$ and $E_{19}$ spaces at different values of $\lambda=10^{-9}\sim10^{-2}$.
Explicitly,
at $\lambda=10^{-9}$, 
the bootstrap matrices $\bm{\mathcal{M}}_9$ and $\bm{\mathcal{M}}_{19}$ are positive semidefinite for
$-2.21900225\ \text{MeV}\leq E_9\leq-2.21900217\ \text{MeV}$
and $-2.22118726\ \text{MeV}\leq E_{19}\leq -2.22118714\ \text{MeV}$,
which are in excellent agreement with the exact results $E_9^\text{exact}$ and $E_{19}^\text{exact}$.
Compared to Figs.~\ref{FigBS}(a) and \ref{FigBS}(b),
it is straightforward to see that 
the allowed $E_{N_\text{max}}$ space shrinks significantly for the same $\lambda$ as $N_\text{max}$ grows.
For example, at $\lambda=10^{-2}$, the allowed $E_{N_\text{max}}$ spaces in the four cases are given by 
$-2.138857\ \text{MeV}\leq E_2\leq-1.610264\ \text{MeV}$,
$-2.220264\ \text{MeV}\leq E_3\leq-1.6758798\ \text{MeV}$,
$-2.226375\ \text{MeV}\leq E_9\leq-2.206636\ \text{MeV}$,
and
$-2.221545\ \text{MeV}\leq E_{19}\leq-2.220773\ \text{MeV}$,
from which one can see that
the sizes of the allowed spaces decrease by about three orders of magnitude as $N_\text{max}$ goes from $2$ to $19$.
This is believed to be closely related to the growing numbers of bootstrap constraints at large $N_\text{max}$s,
which impose more restricted constraints than those at small $N_\text{max}$s.

\section{Conclusions}
\label{Concl}

The idea of bootstrap aims to solve quantum physics problems by exploiting fundamental physical principles and self-consistency conditions.
It was initially proposed in the 1960s as a novel framework to understand the strong interaction
and has been ignored for decades in the era of QCD.
Its real power in solving important physical problems has been appreciated only recently
thanks to noticeable advances in conformal field theories and relativistic scattering amplitudes.
Inspired by recent bootstrap studies in nonrelativistic quantum mechanics,
we report the first bootstrap results in low-energy nuclear physics
by studying the deuteron.
We exploit systematically one of the most fundamental quantum mechanical requirements 
that probability should never be negative.
Combined with other self-consistency conditions,
it is found to be able to determine 
the deuteron energy from the $\slashed{\pi}$EFT Hamiltonian in harmonic oscillator space,
giving numerical results in excellent agreement with exact solutions
from, e.g., direct matrix diagonalization.
Compared with traditional quantum few-body methods,
quantum mechanical bootstrap method is especially prized for the conceptual advantage
to make manifest the connections between the outcomes of quantum theories and the underlying 
physical principles,
which are often hidden behind technical details in traditional methods.
Our study shows that, 
besides simple problems in introductory textbooks on quantum mechanics,
the bootstrap method can indeed be helpful for understanding low-energy properties of nuclear systems in the real world.

\begin{acknowledgments}

This work is supported by the National Natural Science Foundation of China (Grants No.\ 11905103, No.\ 11947211, and No.\ 12035011).


\end{acknowledgments}


\appendix

\end{document}